\documentclass{article}
\usepackage[utf8]{inputenc}
\usepackage{amsmath}
\usepackage{accsupp}
\usepackage{amssymb}
\usepackage{bm}
\usepackage{graphicx}
\usepackage{siunitx}
\usepackage[font=small,labelfont=bf]{caption}
\usepackage{biblatex} 
\graphicspath{ {./images/} }
\begin{document}
\title{The NOSTRA model: coherent estimation of infection sources in the case of possible nosocomial transmission}
\author{David J Pascall, Chris Jackson, \\
Stephanie Evans, Theodore Gouliouris, Chris Illingworth,\\ Stefan Piatek, Julie V Robotham, Oliver Stirrup, Ben Warne,\\ Judith Breuer,
Daniela De Angelis}
\date{}

\maketitle
\section{Abstract}
Nosocomial infections have important consequences for patients and hospital staff: they worsen patient outcomes and their management stresses already overburdened health systems. Accurate judgements of whether an infection is nosocomial helps staff make appropriate choices to protect other patients within the hospital. Nosocomiality cannot be properly assessed without considering whether the infected patient came into contact with high risk potential infectors within the hospital. We developed a Bayesian model that integrates epidemiological, contact and pathogen genetic data to determine how likely an infection is to be nosocomial and the probability of given infection candidates being the source of the infection. 
\section{Introduction}
Nosocomial infections are an important issue facing health systems across the world, impacting both patient survival \cite{goto2013,kraker2011} and their willingness to access healthcare \cite{saah2021}. However, when an infection is found in a patient in a healthcare setting it is often unclear whether the infection was genuinely acquired in hospital or just detected there. These two scenarios have very different implications for decision making. If the infection is the result of an outbreak on a ward, then measures such as ward closures are commonly used to stop further within hospital transmission \cite{wong2015}. This has downstream impacts on both patient treatment and staff time, and thus the economics of the system within which the hospital operates. Given this, methods for assessing both whether an infection is nosocomial, and if it is, whether it is part of a large hospital outbreak, are necessary for coherent decision making in infection control, with knowledge that an infection was not nosocomial allowing the avoidance of ward closures.

Historically, whether an infection was hospital acquired was commonly assessed by the application of heuristic criteria. For example, England's public health agency, the UK Health Security Agency (then Public Health England), in its last major study of nosocomial infections used the their main definition of a hospital acquired infection as being an infection (of the type they were interested in) where symptom onset occurs on at least the 3rd day post-admission \cite{PHE2016}. Moving beyond this heuristic approach to one based on specific evidence related to characteristics of the pathogen of interest and host population would be more principled, with integrating larger amounts of data on a coherent fashion hopefully leading to more accurate assessments. 

Tools have been developed explicitly for the assessment of nosocomiality \cite{stirrup2021,gomezvallejo2016,cohen2004} or the related task of inferring transmission histories (reviewed in \cite{duault2022}). However, these tools cannot simultaneously answer the questions of "Is the infection nosocomial?" and "What is the most likely source?". Treating both questions simultaneously should lead to systematically different answers than would be the case if attempting to answer them individually, because the presence or absence of high probability infection candidates within the hospital should impact our assessment of the probability of whether the infection is nosocomial. Given that the appropriate infection control response depends on whether an infection truly is nosocomial, there is an urgent need for the development of tools that are capable of simulateously answering both of these questions, which can be used routinely by hospital staff in real time to guide management decisions.

Unfortunately, attributing nosocomiality and assessing infection sources of a given infection is a non-trivial task, as generally not all infection candidates will be identified, and even in cases where every possible infection candidate is known, data may only be available for a subset. Hence, any method to approach these problems together (or, indeed, individually) must decide what assumptions are going to be made about this problem of missing infection sources (and thus associated data). This is done by either making assumptions about the nature of the missing data (commonly that data that is available is representative of the missing data) as in \cite{stirrup2021}, or by reducing focus to a circumscribed question that is answerable with just the data observed as in \cite{illingworth2021, illingworth2022, didelot2021}.

One model that directly attempted to assess nosocomiality was HOCI \cite{stirrup2021}. This Bayesian model was developed at the height of the SARS-CoV-2 pandemic to integrate epidemiological and genetic data to attempt to assess whether detected SARS-CoV-2 infections were hospital acquired. To model the epidemiological information, it used a distribution from infection time to symptom onset to assess whether infection was more likely in the hospital or the community given the observed admission and onset times. The genetics were modelled by testing the consistency of the viral isolate from the patient of interest to sequences from the hospital relative to those from the community. This tool was rolled out across NHS hospitals in the UK for real time use by infection control teams in a large study assessing the impact of sequencing on clinical outcomes \cite{stirrup2022}. HOCI, however, did not attempt to make any assessments of the source of the infection if it was determined to have a high posterior probability of being nosocomial. 

In contrast, the A2B model \cite{illingworth2022}, focuses exclusively on source attribution. That is, conditional on there having been an infection in hospital, which individuals are consistent with having been the source. A2B works within a frequentist framework providing p-values on whether the data are more extreme than would be observed under the null model of infection from individual A to individual B, hence the name. Like HOCI, A2B makes use of infection times and sequence information, but it also uses information about the locations of individuals within the hospital, in order to rule out some transmission events.

Here, we, a group of authors comprising many of the original developers of the HOCI and A2B models, present a novel Bayesian model, designed as a conceptual integration of these two models \cite{illingworth2022, stirrup2021}, that we call NOSTRA (NOSocomial TRansmission Assessment), a name inspired by the historical prognosticator Nostradamus. The aims of our model are twofold: 

\begin{itemize}
    \item To provide both a probability a detected infection was acquired within the hospital, and probabilities for the source being within a set of given candidate individuals
    \item To have a low enough runtime to be usable on wards in real-time for clinical decision making 
\end{itemize}

We illustrate the outputs of this model using previously published data \cite{illingworth2021} collected during the early stages of the COVID-19 pandemic at Cambridge University Hospitals NHS Foundation Trust (CUH).

\section{Data and methods}
\subsection{Data}
The data we use to illustrate our model outputs are fully explained in Illingworth et al (2021) \cite{illingworth2021}, but we briefly re-describe it here. The data was collected during prospective COVID-19 surveillance at CUH between the 22nd March and 14th June 2020. Patients were tested for COVID-19 through targeted patient screening in wards with detected hospital-onset outbreaks. The original data comprised five wards, but we focus only on the ward identified as A in Illingworth et al (2021). Due to a large outbreak on this ward, all individuals on the ward were eventually tested, including those exhibiting no symptoms. Final case sets were generated manually by seeking for possible links in a social network diagram generated in FoodChain-lab \cite{weiser2016}.

Patient locations on each day and the date of onset of symptoms (or in the case of aymptomatic infections date of detection) were extracted from the hospital's electronic records. Viral sequences were generated from isolates using the modified ARTIC v2 protocol \cite{quick2020}.

For this study, we performed some post-processing of the genetic data, firstly reducing the alignment for each pair of individuals to those columns which contain no ambiguities or gaps, recorded the length of this reduced alignment, and calculated the number of SNPs between each pair in the reduced alignment.

\subsection{Model}
The goal of the model is to estimate the most likely source of infection for an individual, who we label $B$, whose infection was discovered in hospital using: data on $B$'s movements; the genetic sequence of their infecting pathogen; their times of admission to hospital and symptom onset; and the movements and onset times of a set of $n$ candidate infector individuals in the hospital, who we label $A_1$ to $A_n$. We partition the possible sources of infection into $n+2$ mutually exclusive groups as follows: the first $n$ sources are the candidate individuals, the $A$s; the $n+1$th source, labelled $H$, is any source of infection from the hospital other than these individuals, including visitors; and the $n+2$th source represents infections outside the hospital, and is labelled $C$. 

We set this up as a Bayesian inference problem.  The unknown source of infection $S$ is a categorical variable with $n + 2$ potential values $s  \in \{A_1,...,A_n,H,C\}$.  The goal is to estimate the posterior probability distribution of $S$, $P(S|X)=P(A_1,...,A_n,H,C|X)$. That is, the probability that $B$'s infection came from each of the $n+2$ sources, given some prior distribution for $S$, $P(S)$, and the observed data. All notation is defined in Table 1.

\begin{center}
\begin{tabular}{ | l | p{10cm} | }
  \hline			
  \textbf{Notation} & \textbf{Definition} \\
  \hline
  $B$ & The focal individual of the study whose source of infection is to be determined \\
  $S$ & The set of $n+2$ infection sources \\
  $A_z$ & The infection source consisting of the $z$th candidate individual within the hospital for the infection of $B$ \\
  $H$ & The infection source for the infection of $B$ consisting of all unknown individuals in the hospital, including visitors \\
  $C$ & The infection source for the infection of $B$ consisting of all individuals outside of the hospital \\
  $X$ & All available data \\
  $X_{A_z}$ & The portion of the data containing the symptom onset time of $A_z$, the distance in terms of single nucleotide polymorphisms (SNPs) between their pathogen genomes (assumed to be generated from an alignment with no gaps or ambiguities), and the set of days that $A_z$ and $B$ were at the in the same location \\
  $X_{B}$ & The portion of the data containing the hospital admission time of $B$, and the symptom onset time of $B$ \\
  $L(X|Z)$ & The likelihood of data $X$ given parameters $Z$ \\
  $P(H=h|Z)$ & The conditional probability or conditional probability density of the realisation $h$ of the random variable $H$ given $Z$ \\
  Distribution$(X|Z)$ & The probability or probability density of data $X$ under the given distribution with parameters $Z$ \\
  $T_o^{I}$, $t_o^{I}$  & The random variable describing the symptom onset time of individual $I$, and its realisation, respectively  \\
  $T_i^{I}$, $t_i^{I}$  & The random variable describing the infection time of individual $I$, and its realisation, respectively  \\
  $T_w^{I}$, $t_w^{I}$  & The random variable describing time between infection and symptom onset of individual $I$, and its realisation, respectively  \\
  $T^{I,J}_{MRCA}$  & The random variable describing time in viral generations since the viral isolates from individuals $I$ and $J$ shared a common ancestor \\
  $t_s^I$ & The sampling time of $I$'s pathogen isolate \\
  $t_d^{I,J}$ & The absolute time difference between the sampling time of $I$'s pathogen isolate and $J$'s pathogen isolate, that is $|t_s^I-t_s^J|$ \\
  $t_a^I$ & The hospital admission time of individual $I$ \\
  $t_{\emptyset}$ & The start time of the epidemic \\
  $\Delta^{I,J}$, $\delta^{I,J}$  & The random variable describing the number of SNPs between the viral isolates from individuals $I$ and $J$, and its realisation \\
  $D^{I,J}$, $d^{I,J}$  & The random variable describing whether individuals $I$ and $J$ were in contact on a set of days, and its realisation \\
  $E$ & The per base error probability of the sequencing technology employed to generate the pathogen genome sequences \\
  $G$ & The genome length of the pathogen under study \\
  $G^{I,J}$ & The effective genome length of the alignment between individuals $I$ and $J$ after removing gaps and ambiguities in the alignment \\
  $N_e$ & The effective population size of the pathogen under study at the time of sampling \\
  $M$ & The evolutionary rate of the pathogen under study, in mutations per unit time \\
  $g$ & The mean generation time of the pathogen under study \\
  \hline  
\end{tabular}
\end{center}

We partition the data $X$ into several components, such that $X = \{X_{A_1},...,X_{A_n},X_H\}$. $X_{A_z}$ consists of the onset time of $A_z$, $t_o^{A_z}$, the distance in terms of single nucleotide polymorphisms (SNPs) between their pathogen genomes (assumed to be generated from an alignment with no gaps or ambiguities), $\delta^{A_z,B}$, and whether $A_z$ and $B$ were in the in the same location on each day, $d^{A_z,B}$. $X_H$ consists of the admission time of $B$, $t_a^B$, and the onset time of $B$, $t_o^B$.

\subsection{Bayesian analysis: Prior}
Bayesian analysis requires a prior over $S$. There are multiple ways that this prior could be justified. We use a uniform prior over $S$ for our illustrations with the CUH data.

\subsection{Bayesian analysis: Likelihood}
For Bayesian estimation of $P(S|X)$, we need to define the likelihood of the data $X$ for each potential value (or "hypothesis") for the unknown $S$.  We denote this $L(X|S = s)$, and we now define it in turn for each $S$. 

There are two classes of hypothesis; when $S \in \{H,C\}$ and when $S \in \{A_1,...,A_n\}$. Within each class of hypothesis, the likelihood has the same general structure. We will treat them one at a time.
\subsubsection{Likelihood of the data given that infection was from a non-candidate in the hospital or in the community, $L(X|S\in\{H,C\})$}
Under this class of hypothesis, the infection occurred in the community or from an unknown individual in the hospital.

We make the strong simplifying assumption that each of the components of $X$ are generated independently of one another, and hence the likelihood factorises. This assumption rules out indirect transmission between candidate individuals and the focal individual, as in the scenario $B$'s infection came indirectly from $A_z$ via a person in $H$ or $C$, the onset time of $A_z$ would not be independent of that of $B$. 

Under this assumption:
\begin{equation}
L(X|S=H)=L(X_H|S=H)\prod_{z=1}^{n}L(X_{A_z}|S=H) 
\end{equation}
and
\begin{equation}
L(X|S=C)=L(X_H|S=C)\prod_{z=1}^{n}L(X_{A_z}|S=C)
\end{equation}

We will now derive each component of these likelihoods separately.
\subsubsection{Likelihood of $X_H$ given that infection was in the community, $L(X_H|S=C)$}
This is the likelihood for the onset time of person B, $t_o^B$, given their admission time, $t_a^B$. This is obtained by specifying parametric models for B's (unknown) infection time $T_i^B$, assumed to have density $f()$ , and the ``waiting'' time between B's infection and onset, $T_w^B = T_o^B - T_i^B$, assumed to have density $g()$.  

Suppose we knew the infection time was $t_i^B$, then the onset time is $t_o^B = t_i^B + t_w^B$. The probability of observing $t_o^B$ could then be obtained directly from the model for $T_w^B$, that is, $P(T_w^B = t_o^B - t_b^B)$. However, we do not know the infection time, so the likelihood for $t_o^B$ is determined by integrating this probability over the range of values of $t_i^B$ compatible with having acquired the infection outside of hospital, that is, infection times between the start of the epidemic $t_{\emptyset}$ and the admission time $t_a^B$:
\begin{equation}
  L(X|S=C)=P(T_o^B = t_o^B | S=C) = \int_{t_{\emptyset}}^{t_a^B}  P(T_w^B = t_o^B - t_i^B) P(T_i^B=t_i^B) dt_i^B 
\end{equation}

The admission dates were not collected for the CUH data that we use to illustrate the method, so we set $P(T_o^B = t_o^B |S\in\{H,C\})=1$ for this data.
\subsubsection{Likelihood of $X_H$ given that infection was from a non-candidate in the hospital, $L(X_H|S=H)$}
This is as $L(X_H|S=C)$, except that the integral is taken over the range of infection times that are compatible with infection being acquired from an unidentified individual within the hospital between $t_a^B$ and $t_o^B$.
\subsubsection{Likelihood of $X_{A_z}$ given that infection was from a non-candidate in the hospital or in the community, $L(X_{A_z}|S\in\{C,H\})$}
$X_{A_z}$ contains the onset time of $A_z$, $t_o^{A_z}$, the difference in terms of SNPs between the sequenced genomes of the pathogens infecting $A_z$ and $B$, $\delta^{A_z,B}$, a vector of 1s and 0s describing on which $A_z$ and $B$ were in the same location, $d^{A_z,B}$, and for any unobserved elements of $d^{A_z,B}$ elicited probabilities that $A_z$ and $B$ were in contact on those days, $w(A_z,B)$. Note that, as none of these data are impacted by whether $B$ was infected in the hospital or the community, the likelihood is identical under both $S=H$ and $S=C$. Under these hypotheses, $A_z$ was not the source of $B$'s infection, hence, here we assume that the components of $A_z$'s data ($A_z$'s onset time, the genetic difference between the viruses affecting $A_z$ and $B$, and $A_z$'s co-location with $B$) are independent of each other. As before, this independence is plausible if there was no intermediate transmission between $A_z$ and $B$. This implies that  $A_z$ did not infect $B$ (and vice versa), their co-locations, $d^{A_z,B}$, are independent of the other data in $X_{A_z}$ and the genetic data is independent of the epidemiological data and thus can be modelled separately.

Given this, we have:

\begin{align}
\begin{split}
  L(X_{A_z}|S\in\{C,H\})&=P(\Delta^{A_z,B}=\delta^{A_z,B},T_o^{A_z}=t_o^{A_z},  D^{A_z,B}=d^{A_z,B}|S\in\{C,H\}) \\
  &=P(\Delta^{A_z,B}=\delta^{A_z,B}|S\in\{C,H\})P(T_o^{A_z}=t_o^{A_z}|S\in\{C,H\})\\
  &P(D^{A_z,B}=d^{A_z,B}|S\in\{C,H\}) 
\end{split}
\end{align}

where $\Delta^{A_z,B}$ and $D^{A_z,B}$ are the random variables underlying the observed genetic data $\delta^{A_z,B}$ and co-location data $d^{A_z,B}$. 

We already have a parametric model for onset time, which we applied to $B$'s onset above. The same model can be applied to $A_z$'s onset time, with the integration for $t_i^{A_z}$ being over the possible infection dates for $A_z$, that is between $t_\emptyset$ and $t_o^{A_z}$.

Because of there is no long a dependence on the admission time, we could include this in our illustration of the method using the CUH data. For simplicity we use a uniform distribution between $t_{\emptyset}$ and $t_o^{A_z}$ for the the distribution of $T_i^B$. Following, Illingworth et al. 2022 \cite{illingworth2022,he2020} we give $T_w^B$ a lognormal distribution with mean 1.434 and standard deviation 0.6612. We set $t_{\emptyset}$ to 1st January 2020.

For the co-locations, $d^{A_z,B}$, we follow the the approach taken in A2B \cite{illingworth2022}. Individuals either are or are not in contact on any particular day, giving a total of $2^{|D|}$ potential contact history vectors for $|D|$ days. Assuming that none of these contact histories are more or less likely than any other given $A_z$ did not transmit to $B$, the observed contact history then has probability $P(D^{A_z,B}=d^{A_z,B}|S\in\{C,H\})=0.5^{|D|}$.

To specify $P(\Delta^{A_z,B}=\delta^{A_z,B}|S\in\{C,H\})$ we make use of the coalescent \cite{kingman1982b, kingman1982a, tajima1983}. All the necessary genealogical theory for this section is reviewed in Hudson 1990 \cite{hudson1990}. Assume that the viruses are evolving under a Poisson process with rate, $M$. Under the coalescent, a.k.a. the Wright-Fisher model in the limit of infinite population size, the number of generations to the most recent common ancestor (MRCA), for two randomly chosen individuals, is exponentially distributed with rate given by the inverse of the (effective) population size, $N_e$. Let $T^{A_z,B}_{MRCA}$ represent this random variable. Since this represents the number of generations since the pathogens infecting $A_z$ and $B$ last shared a common ancestor, the pathogens are separated by $2T^{A_z,B}_{MRCA}$ generations of independent evolution at rate $M$. Hence, during the time in standard units spanned by these generations, we would expect the the number of SNPs generated through evolution to be Poisson distributed with mean $2T^{A_z,B}_{MRCA}MgG$. 

Our assumption that the alignment has no gaps or ambiguities is unrealistic, so we create an accounting variable for each candidate individual and the focal individual $G^{A_z,B}$, which corresponds to the effective genome size after ambiguous sites and gaps have been removed. We assume that the alignment of $G^{A_z,B}$ length is comparable to the unrealised complete alignment of length $G$. Note that, in all cases, $G^{A_z,B} \leq G$. We use this new variable to correct the mean to $2T^{A_z,B}_{MRCA}MgG^{A_z,B}$.

As $T^{A_z,B}_{MRCA}$ is an exponentially distributed random variable with rate $1/N_e$, $2T^{A_z,B}_{MRCA}MgG^{A_z,B}$ is also exponentially distributed with rate, $\frac{1}{2MgN_eG^{A_z,B}}$. As a Poisson distribution with a Gamma-distributed random rate parameter is equivalent to a Negative Binomial distribution, the number of mutations generated through evolution between the two sequences can be modelled as NB($r=1,p=\frac{1}{1+2MgN_eG^{A_z,B}}$).

In addition, there would then be differences added by sequencing error in both genomes. We can model the number of sequencing errors as a Binomial random variable with probability $E$, the per base error probability and number of trials $2G$, double the genome size, as this occurs in both genomes. As $E$ will be small and $2G$ is large, we approximate this Binomial distribution with a Poisson distribution with rate $2EG$. Again, we correct for the observed genome length by modifying this rate to $2EG^{A_z,B}$.

Therefore the total number of genetic differences between the two genomes is the sum of the Negative Binomial distribution describing the SNPs generated though mutation and the Poisson approximation to the Binomial distribution describing the SNPs generated through sequencing error (assuming no back mutation). The sum of a Negative Binomial distributed random variable and a Poisson distributed random variable is Delaporte distributed \cite{schroter1990}. Hence, the likelihood is:

\begin{multline}
P(\Delta^{A_z,B}=\delta^{A_z,B}|S \neq A_z) =\\
\text{Delaporte}\left(\delta^{A_z,B}|\alpha=2MgN_eG^{A_z,B},\beta = 1,\lambda = 2EG^{A_z,B}\right)
\end{multline}

Note that if the isolates are collected on different days with time difference in standard units, $t_d^{A_z,B}$, the distribution of the time between them would be Exp($\frac{1}{2gN_e}$)$+ t_d^{A_z,B}$ instead of just Exp($\frac{1}{2gN_e}$). This can be accounted for by modifying the $\lambda$ parameter of the Delaporte distribution from $\lambda=2EG^{A_z,B}$ to $\lambda=G^{A_z,B}\left(2E+\frac{t_d^{A_z,B}M}{G}\right)$.

For our analysis of the CUH data, we set $N_e$ to 51, $g$ to 5.5, $M$ to \num{1.829e-6}, and we follow Illingworth et al. 2022 \cite{illingworth2022} in approximating $2EG^{A_z,B}$ with 0.404.
\subsubsection{Likelihood of the data given that infection was from a candidate individual, $L(X|S\in\{A_1,...,A_n\})$}
Under this class of hypothesis, $B$ was infected by one of the candidate individuals $A_z$. We assume then that the data $X_{A_j}$, that describe the relationship between $B$ and each other individual $A_j$, $j \in \{1,...,n \setminus z\}$, are independent between each $A_j$, and independent of the data $X_{A_z}$ that describe the relationship between the infecting $A_z$ and $B$. That is:
\begin{equation}
    L(X|S=A_z)=L(X_H,X_{A_z}|S=A_z)\prod_{j=1, j\ne z}^{n}L(X_{A_j}|S=A_z)
\end{equation}
$L(X_{A_j}|S=A_z)$ takes the same form as $L(X_{A_z}|S\in\{C,H\})$. All that remains is to generate a parametric model for $L(X_H,X_{A_z}|S=A_z)$.
\subsubsection{Likelihood of $X_H$ and $X_{A_z}$ given that infection was from the candidate individual $A_z$, $L(X_H,X_{A_z}|S=A_z)$}
We partition the data into event times $T$, genetic distance $\Delta$, and co-location $D$ components, and rearrange in terms of the conditional probability of $B$'s data given $A_z$'s onset time $T_o^{A_z}$. 
\begin{align}
\begin{split}
L(X_H,X_{A_z}|S=A_z)&=P(\Delta^{A_z,B}=\delta^{A_z,B},T_o^{A_z}=t_o^{A_z},T_o^B=t_o^B, D^{A_z,B}=d^{A_z,B}|S=A_z)\\
&=P(\Delta^{A_z,B}=\delta^{A_z,B},T_o^B=t_o^B,D^{A_z,B}=d^{A_z,B}|T_o^{A_z}=t_o^{A_z},S=A_z)\\
&P(T_o^{A_z}=t_o^{A_z}|S=A_z)
\end{split}
\end{align}
$P(T_o^{A_z}=t_o^{A_z}|S=A_z)$ is provided by the model in 3.3.4. 

We then obtain $P(\Delta^{A_z,B}=\delta^{A_z,B},T_o^B=t_o^B, D^{A_z,B}=d^{A_z,B}|T_o^{A_z}=t_o^{A_z},S=A_z)$ using the same technique as in the A2B model \cite{illingworth2022}.  This involves expanding this term by summing it over $B$'s unknown time of infection $T_i^B$, and assuming that $B$'s onset time, the genetic distance of $B$'s  pathogen from $A_z$'s, and $B$'s co-location with $A_z$ are conditionally independent given this infection time.  Specifically: 

\begin{align}
\begin{split}
 P(\Delta^{A_z,B}=\delta^{A_z,B},T_o^B=t_o^B, D^{A_z,B}=d^{A_z,B}|T_o^{A_z}=t_o^{A_z},S=A_z) =\\
 \sum_{t = t_\emptyset}^{t_o^B} P(T_i^B = t|T_o^{A_z}=t_o^{A_z},S=A_z) P(T_o^B = t_o^B|T_i^B = t,T_o^{A_z}=t_o^{A_z},S=A_z)\\
 P(\Delta^{A_z,B}=\delta^{A_z,B}|T_i^B = t,T_o^{A_z}=t_o^{A_z},S=A_z)\\
 P(D^{A_z,B}=d^{A_z,B}|T_i^B = t,T_o^{A_z}=t_o^{A_z},S=A_z)
\end{split}
\end{align}

Furthermore, we note that the onset time of $A_z$ provides no extra information after conditioning on the infection time of $B$, so we can simplify as follows:
\begin{align}
P(T_o^B = t_o^B|T_i^B = t,T_o^{A_z}=t_o^{A_z},S=A_z)&=P(T_o^B = t_o^B|T_i^B = t,S=A_z)\\ P(\Delta^{A_z,B}=\delta^{A_z,B}|T_i^B = t,T_o^{A_z}=t_o^{A_z},S=A_z)&=P(\Delta^{A_z,B}=\delta^{A_z,B}|T_i^B = t,S=A_z)\\ 
P(D^{A_z,B}=d^{A_z,B}|T_i^B = t,T_o^{A_z}=t_o^{A_z},S=A_z)&=P(D^{A_z,B}=d^{A_z,B}|T_i^B = t,S=A_z)
\end{align}

Any distributional form could be assumed for $P(T_o^B = t_o^B|T_i^B = t,S=A_z)$, and this choice should be specific to the pathogen in question and informed from its epidemiological literature.

For the term $P(\Delta^{A_z,B}=\delta^{A_z,B}|T_i^B = t,S=A_z)$, we take a similar the approach for the genetics used for $L(X_{A_z}|S\in\{C,H\})$, but the situation is drastically simplified as the infection time is known. We make the assumption of no within host variation in the pathogen, so that there is no risk of incomplete lineage sorting and the time of the MRCA of $B$ and $A_z$, is exactly the infection time of $B$. If $t$ is the infection time of $B$, and $t_s^B$ and $t_s^{A_z}$ are the sampling times of the pathogen genomes of $B$ and $A_z$ respectively, then there has been $|t_s^B-t|+|t_s^{A_z}-t|$ time units of of independent evolution for the pathogens. Given the Poisson process assumption for mutation being used, we expect the number of SNPs between the genomes to follow a Poisson distribution with mean $(|t_s^B-t|+|t_s^{A_z}-t|)MG$, which after accounting for partial observation, becomes $(|t_s^B-t|+|t_s^{A_z}-t|)MG^{A_z,B}$. Again, we assume that additional mutations are generated by sequencing error, following a Poisson distribution with mean $2EG^{A_z,B}$. As the sum of two Poisson random variables is Poisson, this gives that:

\begin{multline}
P(\Delta^{A_z,B}=\delta^{A_z,B}|T_i^B = t,S=A_z)=\\
\text{Poisson}\left(\delta^{A_z,B}|\lambda=(|t_s^B-t|+|t_s^{A_z}-t|)MG^{A_z,B}+2EG^{A_z,B}\right)
\end{multline}

Finally, consider the likelihood term $P(D^{A_z,B}=d^{A_z,B}|T_i^B = t,S=A_z)$ for the history of co-location $d^{A_z,B}$ of individuals $A_z$ and $B$ under a hypothesis that $A_z$ infected $B$ at time $t$.  These data consists of a vector of $d^{A_z, B}$, where $d^{A_z,B}_c = 1$ or $d^{A_z,B}_c = 0$ if $A_z$ and $B$ are known to have been in co-located, or not in co-located, respectively, for each day $c$.  On some occasions $d^{A_z,B}_c$ is unknown, and we assume we have an elicited probability $w_c(A_z,B)$ that they were in contact then. In the case of fully observed location data, $w(A_z,B)$ is undefined.

Then, assuming conditional independence of the co-location data on each day, we have
\begin{equation}
    P(D^{A_z,B}=d^{A_z,B}|T_i^B = t,S=A_z) = \prod_c  p_c(A_z,B,t)
\end{equation}
\noindent say, where

\begin{equation}
p_c(A_z,B,t) = P(D^{A_z,B}_c=d^{A_z,B}_c|T_i^B = t,S=A_z)
\end{equation}

\noindent at times $c$ when $d^{A_z,B}_c$ is observed. When contact status is unobserved, this likelihood contribution is defined as a weighted average over the missing contact indicator $d^{A_z,B}_c$, using the estimated contact probabilities as weights, giving: 
\begin{align}
\begin{split}
  p_c(A_z,B,t) = & (1 - w_c(A_z,B)) P(D^{A_z,B}_c=0|T_i^B = t,S=A_z)  +\\
  & w_c(A_z,B) P(D^{A_z,B}_c=1|T_i^B = t,S=A_z)
\end{split}
\end{align}

To specify these likelihood contributions, firstly note that if $T_i^B = t$, then $A_z$ and $B$ must have been in contact at day $t$, since contact is necessary for infection.  Therefore if $c=t$, then $P(D^{A_z,B}_c=d^{A_z,B}_c|T_i^B = t,S=A_z)=p_c(A_z,B,t)$. At any other time $c$ (following \cite{illingworth2022}) we suppose that all observed co-location patterns are equally plausible, implying that $P(D^{A_z,B}_c=d^{A_z,B}_c|T_i^B = t,S=A_z) = 0.5$. 

In our analysis of the CUH data, we again approximate $2EG^{A_z,B}$ with 0.404.
\subsection{Bayesian Analysis: Inference}
With the likelihood and priors defined, the joint posterior is therefore:

\begin{equation}
    P(S = s|X) = \frac{L(X|S = s)P(S=s)}{\sum_{j \in \{A_1,..,A_n,H,C\}}L(X|S=j)P(S=j)}
\end{equation}

\section{Results}
NOSTRA provides an estimate of the posterior probability for all of the possible infection sources (Fig 1), with rows corresponding to focal individuals and columns to infection sources. The probability of nosocomial infection, shown in the final column, is then the sum of the probabilities over the candidate individuals and hospital component. The estimated posterior probabilities of the hospital, $H$, and community, $C$, are always estimated to be the same in the CUH data we have used here. This is because the data was not collected for this project, and as such, admission times for all patients were unavailable. The admission time is what identifies the hospital and community likelihoods from one another, and thus, given that they had the same prior probability, they end up with the same posterior probability.

As NOSTRA is capable of running with only symptom dates, we can assess the impact of each of the data sources on the posterior by adding one at a time. This is shown in Fig 2 where the change in posterior probability of the infection sources as different data is visible. In this dataset, the genetics has a very large impact on the generated posterior probabilities. This appears to be driven by the genetic consistency of specific candidate individual's viral isolates to the focal individual's viral isolate dramatically reducing the posterior probabilities of the hospital and community compartments. For this dataset, the impact of the location information on the posterior is contingent on the data already in the model, with it only causing a large change in posterior probabilities if the genetics has already been added. This is because the strength of the location data is to rule out transmission by identifying potential transmission pairs who never interacted at the appropriate time and thus are very unlikely to have been linked. Hence, the location data has the largest impact when it removes infection sources that were favoured by the genetics.

\begin{center}
\includegraphics[width=\linewidth]{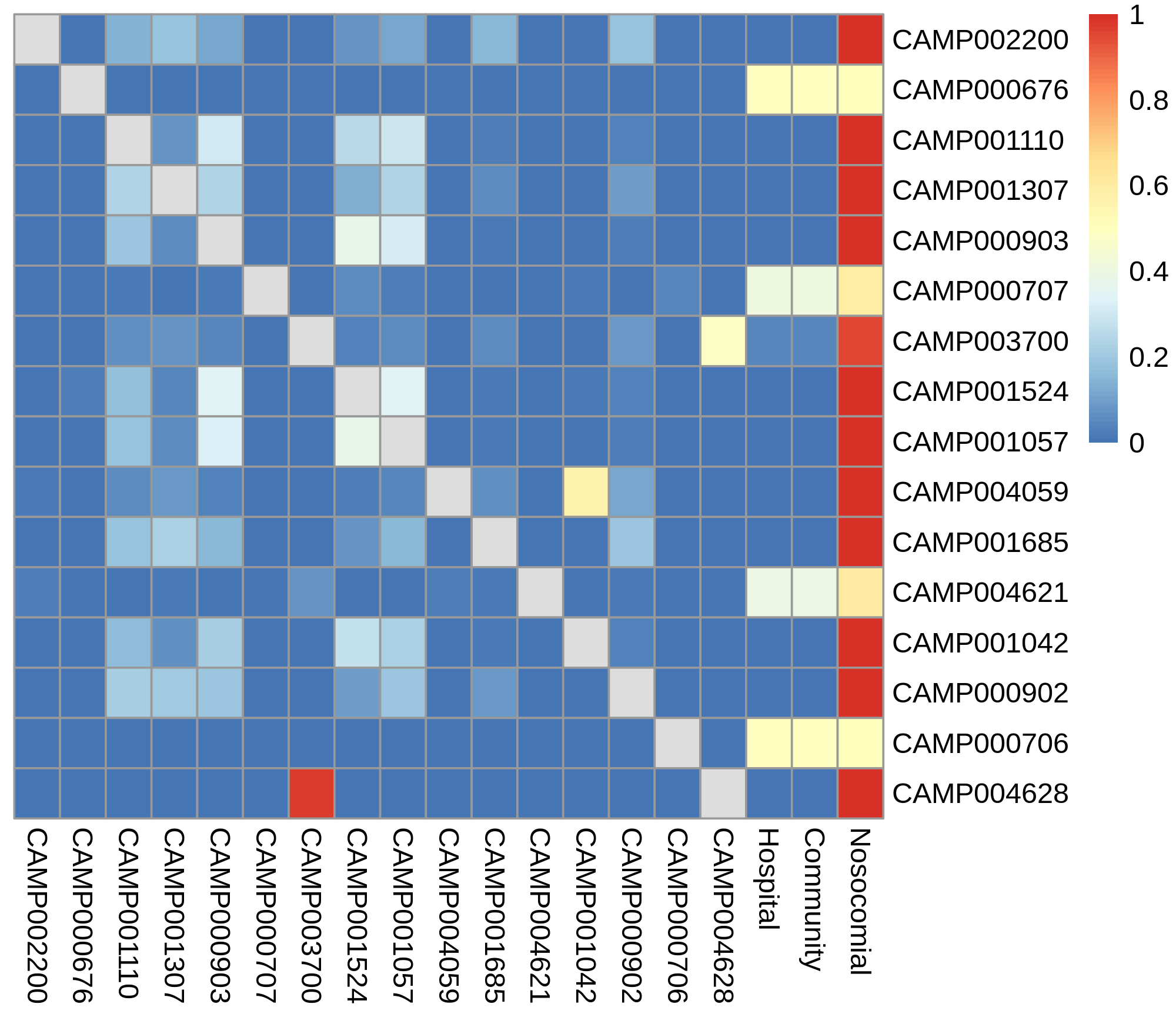}
\captionof{figure}{A visualisation of the output of the model applied to the full CUH Ward A dataset. Each row corresponds to a candidate individual, each column, except the last, to a potential infection source. Cells are coloured by the posterior probability of that infection source. The last column shows the posterior nosocomiality probability.}
\end{center}

\begin{center}
\includegraphics[height=\textheight]{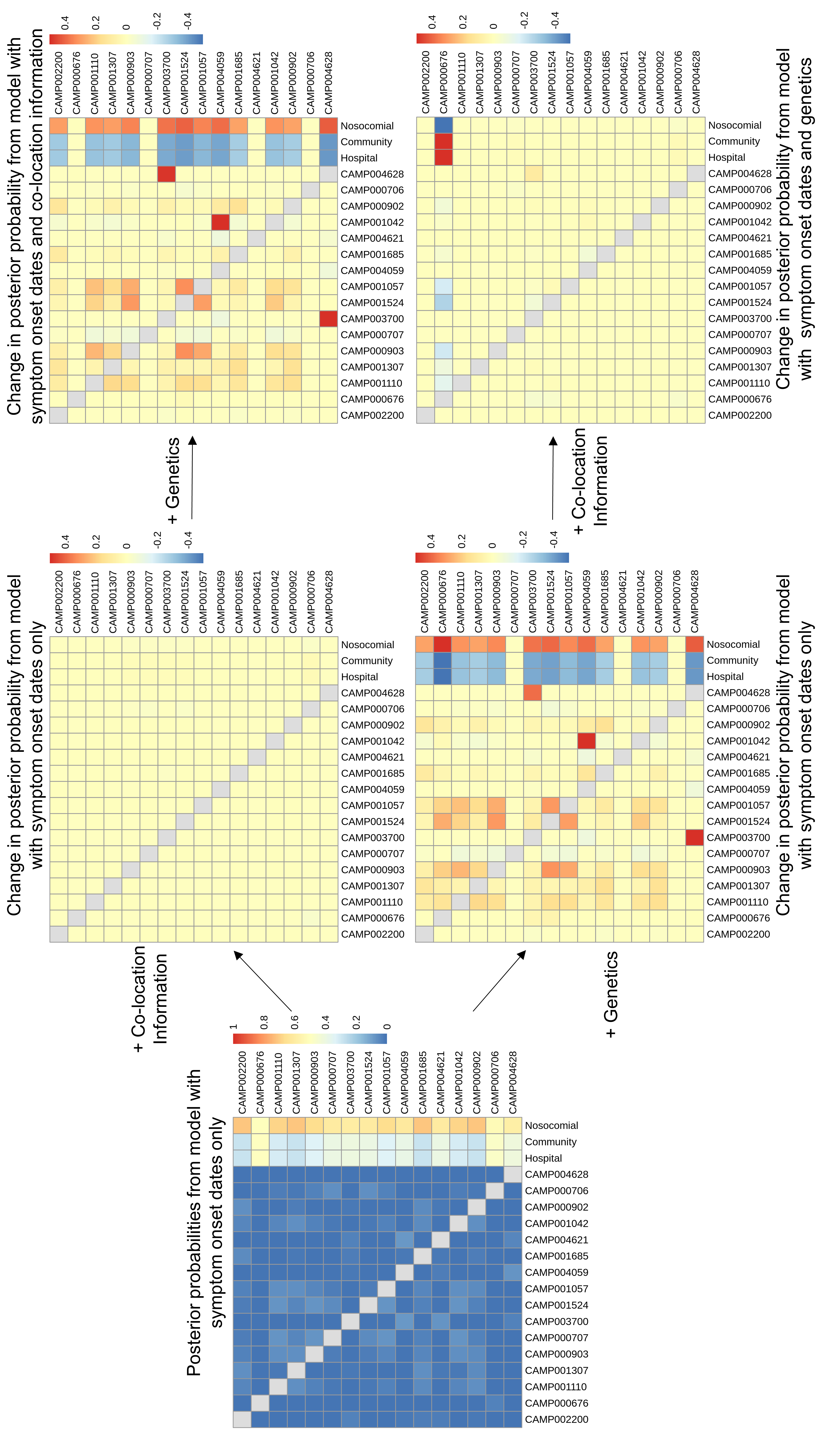}
\captionof{figure}{The impact of adding data sources on the assessed probability of each infection source. The far left panel shows the posterior probabilities when only symptom onsets are provided. Each panel adds a new data source and shows the change in posterior probability with the addition of that data from the panel it is linked to by the arrows.}
\end{center}

\section{Discussion}
We have presented a new model that integrates both epidemiological and genetic data to give a posterior distribution over potential infection sources of an individual. This new model represents the conceptual unification of the HOCI tool \cite{stirrup2021} and the A2B model \cite{illingworth2022}. As all these the terms in our model are mathematically tractable, we get the posterior distribution over sources in analytic form, allowing us to avoid any numerical integration and keep runtime low. There are few published models designed to estimate whether an infection is nosocomial \cite{stirrup2021,gomezvallejo2016,cohen2004}, and to our knowledge none that are designed to jointly estimate the probability of nosocomiality and infection sources within the hospital. Thus, NOSTRA fills an empty niche. 

The results from on the CUH dataset give us some evidence on the kinds of data that may be useful to collect for the assessment of nosocomiality, irrespective of the method that is to be applied. In this case, the genetic data was very informative, as it allowed high probability candidates within the hospital to be identified. This suggests that routine sequencing of hospital pathogens may allow better assessments of nosocomiality, as well as being useful for tracking transmission networks. As mentioned above, admission times were not available for this dataset. Had they have been available, we believe that they would have been the most informative data in the model. This is because the waiting time between infection and onset provides a great deal of information about the location of the infection. If this waiting time is almost always less than five days, and the patient was admitted six days ago, then assessed probability of nosocomialty by NOSTRA is always going to be high, even if there are no genetically consistent identified candidates. Therefore, the genetic and location information are going to have the most impact on the assessed probabilty of nosocomiality in "difficult" cases. That is, when the timing of infection are consistent with both hospital and community transmission, either because the observed time is right in the middle of the waiting time distribution, or because the waiting time is highly variable. Thus, we theorise that hospital sequencing of isolates may be especially valuable for pathogens with highly variable incubation times.

While NOSTRA is currently the only real option for joint estimation of nosocomiality and infection sources, it has some important caveats and limitations that potential users should take into account.

A first caveat regards our handling of the genetics of the pathogen. We implicitly assume that there is a single genetic type at any time in each host. Explicitly we assume that given data for $A_z$ and $B$, when $A_z$ was the infection source, the time of the most recent common ancestor was the point of infection, $t_i^B$. This is only true if $A_z$ had no within host variation in its pathogen. If there is within host variation, this actually represents a lower bound on the time to the common ancestor, due to the potential for incomplete lineage sorting \cite{pamilo1988}. It is unlikely that this is likely to cause a large issue in most cases in which we envision that NOSTRA would be applied, given that the short generation times of most respiratory viruses means that the upper bound on the time to the most recent common ancestor is likely to be close to the lower bound. However, in infections with long generation times, where large amounts of within host diversity may be generated and maintained, this assumption may have a large impact, causing the number of expected SNPs between the infector and infectee to be underestimated. To account for this, a more complicated model allowing for pathogen diversification within hosts after infection would be required. Thus, we advise that users should not use NOSTRA for pathogens with long generation times.

A second caveat is about our handling of missing data. While NOSTRA can run with all data other than the onset times missing, we are making strong assumptions about the nature of that missingness in order to do this. We assume that the data is missing completely at random. That is, that the missing data is a random subset of the full data and it being missing is independent of both the values of observed and unobserved data. The degree to which this assumption holds is likely to depend on the specifics of the pathogen and hospital that this is being applied to. For example, samples with low pathogen load are known to fail sequencing more frequently than those with high load, so in a case where failed sequences are not reattempted, a missing genetic sequence may be indicative of someone early or late in their infection course, or it may simply be that their sample was not sent for sequencing for an unrelated reason. A full understanding of the data providence is necessary to assess whether this assumption is reasonable, and as such whether the model is appropriate to be applied in the case of the user's specific missing data. Another kind of missing data is that of unrecognised infector candidates in the hospital. To some extent our "Hospital" infection source allows for the true infector in the hospital to be unidentified. However, semi-pathological behaviour is likely to occur if the true infector is unidentified, but there is a consistent infector in the set of identified candidates. Under this circumstance, the consistent infector will be likely to be assessed as having high posterior probability at the expense of the "Hospital" source. Given this, the "Hospital" source should be considered a guard against the possibility that there are no likely identified candidates, but the timings of infection strongly suggest nosocomiality.

A third thing that users should take into account when using NOSTRA is its prior. In our analysis of the CUH data, we used a uniform prior over infection sources. This has the disadvantage of placing a very high prior probability on nosocomiality when the number of candidate individuals is high. For this reason, in actual usage, firstly defining a prior probability of nosocomiality, and then distributing the rest of the probability equally over the hospital-associated infection components may be preferable.

The required complexity of NOSTRA to model the multiple data sources that can be inputted to it means that it has higher requirements for prior knowledge about the biology of the pathogen of interest than many other similar epidemiological models. Specifically, studies must have been performed estimating an effective population size for the pathogen in the recent past, in order for the genetic likelihoods to be calculable. This is not a large problem for well studied infections of the kind that we believe NOSTRA is likely be applied to (e.g. COVID-19, RSV and influenza) where phylogenetic studies are regularly being performed and a values will be accessible in the literature. However, it does represent a limitation with respect to understudied or novel pathogens, where effective population size estimates may not be available.

Another potential limitation is that to ensure tractability of the posterior distribution, we had to make strong independence assumptions between the different data types. The assumption of independence between the genetic data and the epidemiological data for $A_z$ and $B$, when $A_z$ was the infection source, being one notable for example. In reality, there will be complex interrelations between these two data types, given that $B$'s epidemiological data and the genetic distance from the isolate from $A_z$ should depend on $A_z$'s epidemiological data through its influence on the infection time of $B$. This could lead to over- or under-estimation of the probability of $A_z$ being the infection source of $B$ depending on the precise combination of the data.

One final limitation relates to the use of the coalescent to model "unrelated" genetic sequences. The form of the coalescent we use makes two assumptions that might be problematic. Firstly, that there is no selection. Over short periods of time, where there is one dominant genetic type, this might be approximately true, but over longer time periods, it will definitely not be. Secondly, that there is no population structure. It is likely that there will be some degree of spatial structuring, and that the sequences in the hospital will be more closely related than would be the case if they were drawn at random from the entire population. This means that $T^{A_z,B}_{MRCA}$ is likely to have a mean that is too high, i.e. that the time to coalescence would be shorter than the would be expected for two sequences drawn at random, and consequently, the expected number of SNPs between the isolates would be overestimated. Both of these issues could potentially be resolved by modifying the form of the coalescent used, likely at the cost of more prior knowledge being required, but that goes beyond the scope of this work.

Despite the above limitations, however, we believe that NOSTRA represents a step forward in data integration for nosocomial infection detection. Our tool provides the probability that an infection is nosocomial as well as the probablity that certain given candidate individuals were the source, something that was not previously available. We have reached the point that there are now multiple models that purport to assess nosocomiality in the literature, but we are limited by the absence of datasets where both the truth is known and the answer is non-trivial, so the accuracy of the assessments cannot be quantified or compared. Finding such datasets should be a focus going forward, so that clinicians and medical statisticians can choose to implement the model that they would expect to perform best for their specific scenarios.

\section{Data}
The code for the analyses and the implementation of the NOSTRA model is available at https://github.com/dpascall/NOSTRA-model.

\section{Funding}
DJP is funded by a NIHR award to JB (NIHR200652). DJP was funded by UKRI through the JUNIPER consortium (MR/V038613/1). DJP, CJ, DA were funded via the MRC Biostatistics Unit Core Award (MC UU 00002/11). JVR was supported by the National Institute for Health and Care Research (NIHR) Health Protection Research Unit (HPRU) in Modelling and Health Economics, which is a partnership between the UK Health Security Agency (UKHSA), Imperial College London, and the London School of Hygiene and Tropical Medicine (NIHR200908). The funders had no input into the design of the study or the decision to publish.

\printbibliography

\end{document}